

\long\def\UN#1{$\underline{{\vphantom{\hbox{#1}}}\smash{\hbox{#1}}}$}
\def\NP{\vfil\eject}
\def\NI{\noindent}

\magnification=\magstep 1
\overfullrule=0pt
\hfuzz=16pt
\voffset=0.0 true in
\vsize=8.8 true in
\baselineskip 20pt
\parskip 6pt
\hoffset=0.1 true in
\hsize=6.3 true in
\nopagenumbers
\pageno=1
\footline={\hfil -- {\folio} -- \hfil}

\hphantom{A}

\hphantom{A} \hfill\hfill Revised October 1994

\vskip 2.0 true in

\centerline{\UN{\bf Series Analysis of Tricritical Behavior:}}
\centerline{\UN{\bf Mean-Field Model and Slicewise Pad\'e Approximants}}

\vskip 0.4in

\centerline{{\bf Joan Adler}$^a$\ \ and\ \ {\bf Vladimir Privman}$^b$}

\vskip 1.0 in

\NI{$^a$}{\sl Department of Physics, Technion --- Israel Institute of
Technology, Haifa 32000, Israel}

\NI{$^b$}{\sl Department of Physics, Clarkson University,
Potsdam, New York 13699--5820, USA}

\vfill\vfill\vfill

\centerline{\bf ABSTRACT}

A mean-field model is proposed as a test case for tricritical series
analyses methods. Derivation of the 50$^{\rm th}$ order series for the
magnetization is reported. As the first application this series is
analyzed by the traditional slicewise Pad\'e approximant method
popular in earlier studies of tricriticality.

\vfill

\NI {\bf PACS numbers:}\  05.50.$+$q,\  64.60.Fr.

\NP

\NI{\bf 1.\ \ INTRODUCTION}

\

Development of algorithms for the  numerical investigation of
tricritical behavior in a two-variable  phase diagram has been an
elusive goal for many years in the context of both magnetism and
polymer studies.  Two such  systems  that have been the focus
of special interest are random field models for which the
tricritical point is expected in the two-parameter space of
temperature and randomness strength [1], and the $\theta$-point
transition of linear polymers [2] for which the second variable
is the ``stickiness'' fugacity leading to collapse. For
these and other systems with tricritical points, all
the standard large-scale numerical methods have been utilized: Monte
Carlo [3,4,5] and transfer matrix techniques [5,6] (the latter for
2D models), and series analyses [7].

There are many complicating factors for such studies. The models
involved are rarely simple and must be analyzed at
multiple points in the two-parameter space.
A ``test problem'' with an exact solution and a series expansion
(such as the 2D Ising model
which is widely used in calibrating new techniques for second order
transitions) has been missing until now for two-variable problems
with tricritical points.

Some of the technical problems are specific to the numerical technique.
In the random field and $\theta$-point transition models
reaching true equilibrium in Monte Carlo simulations, for instance, is
complicated by slowdown effects in the low-temperature or dense
(collapsed) regions of the phase diagram.
We shall not address simulations
{}further in this paper but note that there is considerable controversy
over tricritical behavior between different
recent studies of the 2D collapse transition [4]: $\theta$
vs.~$\theta^\prime$ points, etc.

Series expansions do not suffer from
equilibration problems, and  in many
cases lattice models are amenable to the generation of series for
all points in the two-variable phase diagram. In particular,
two-variable series have been developed for
both Ising random field (15$^{\rm th}$ order in general dimension) [1]
and  several polymer problems [2].
However for series the main complication is that
the very nature of the tricritical
behavior means that techniques for studying both first and second order
transitions must be applied. While excellent techniques for identifying
{}first order transitions via simulation have recently been developed
[8], methods to identify first order transitions
{}from series expansions are unreliable [9]
unless  both low and high
temperature series exist. In one notable
case (the FCC-lattice Blume-Capel model)
where such expansions
on both sides of the transition were developed [10]
a satisfactory characterization of
tricriticality was made from series.

Despite the absence  of  a systematic, well tested, approach
to studying tricriticality when series from only one temperature
direction are available, some attempts to do so  have been made.
The variety of the makeup methods used, frequently
resulted in differences in answers to the basic question concerning the
existence of a tricritical point, not to speak of
exponent and other parameter estimates
which could be attributed to the diversity of methods rather then to
the quality of the series expansions available.

This work reports two developments towards systematizing tricritical
series analyses. Firstly, we derive a test series based on a
mean-field model with a tricritical behavior which is well understood
and has most of the features of ``real'' tricritical points of 2D and
3D systems. Secondly, we apply the standard ``slicewise'' Pad\'e
method to this new series. We identify those features of the Pad\'e
approximant approach which can be regarded as signatures of a
tricritical point in the phase diagram and which were noted in some
early studies of tricriticality by series [7].

We conclude,
however, that this most straightforward Pad\'e method is not suitable
as an accurate and systematic general analyses technique, and it can
be only used for exploratory studies or in conjunction with other
information. Application of the slicewise Pad\'e method to certain
random-field model series will be reported in a forthcoming publication.
However, the door is still wide open for developing a systematic
series-analysis method, possibly based on elaborations of the
two-variable differential approximant techniques used successfully
{}for bicritical points [11].

One interesting aspect of our test series derivation, reported
in Section~2, and its Pad\'e analysis in Section~4, is that both rely
heavily on novel large-scale computational abilities. Series
derivation required a large Mathematica run, whereas Pad\'e analysis
employing simultaneously many Pad\'e-pole calculations and extensive
graphics representations of the data, revealed new features not
accessible to earlier studies from the seventies [7]. Thus,
{}future tricritical series analyses are likely to be large-scale
computational projects. Section~3 summarizes the tricritical
phase diagram of the mean-field model used in test-series
studies. Finally, Section~5 is devoted to some concluding remarks and
to acknowledgments.

\NP

\NI{\bf 2.\ \ THE MEAN-FIELD MODEL}

\

In this section we report derivation of a low-temperature
two-variable series for a tricritical point in an infinite range
model with mean-field critical behavior. There are two kinds of
solvable Ising-type infinite-range models. In the first and more
{}familiar type the spins are $\pm 1$, but their interaction energy
which is the function of the total magnetization, is essentially
arbitrary. Usually it is selected (or Taylor-expanded) as a polynomial
in the total magnetization so that the resulting constrained free
energy, as function of the magnetization, $m$, resembles the Landau
expansion.

The second type of a mean-field model [12] is defined by having
a simple quadratic energy but a complicated entropy-like
contribution due to assigning essentially arbitrary measure in the
evaluation of the partition function of a system of scalar spins
which vary in $(-\infty, +\infty)$.

Both types of infinite-range models suffer from the difficulty
that the low-temperature, $T$, behavior is different from the
short-range lattice models. Indeed, for short-range models the
low-$T$ series expansions are in terms of the Boltzmann factors of
excitations above a reference ground state, of the form
$\exp(-\Delta E / kT)$, where $\Delta E=O(1)$ is the energy cost due
to a local structure (overturned spins, broken bonds, etc.).
However, this ``locality'' of the excitation structure is lost for
infinite-range models. The $\pm 1$-spin models have entropic
contribution to the constrained (fixed-$m$) free energy with
singularities of the type $\sim (1-m) \ln (1-m)$ near magnetization
$m=1$ (and similarly near $m=-1$). Thus setting up a low-$T$ expansion
presents a mathematical challenge.

The models of [12] are less troublesome in this
respect. One can get a well-controlled series
in powers of $T$ itself. This ``soft'' $T$-dependence is an
artifact of the infinite-range model, and there are some other
artificial features near $T=0$, but the series is well
defined and can be derived in closed form to any fixed order given
sufficiently powerful computational facilities. Thus, we choose to
work with the model of [12] here.

The energy of the interacting scalar spins, $\sigma_i$, is taken as

$$ E=-{J \over 2N} \left( \sum\limits_{i=1}^N \sigma_i \right)^2
\;\; , \eqno(2.1)$$

\NI where $N$ is the number of spins, and $J>0$. The partition
{}function is defined as

$$ Z=\int \ldots \int \exp (-E/kT ) \prod\limits_{i=1}^N d \mu \left(
\sigma_i \right) \;\; , \eqno(2.2)$$

\NI where the spins are weighed with measure $d \mu \left( \sigma
\right) $. The order parameter is obtained
{}from

$$ m=Z^{-1} \int \ldots \int \sigma_1 \exp (-E/kT )
\prod\limits_{i=1}^N d \mu \left( \sigma_i \right) \;\; . \eqno(2.3)$$

The Gaussian-integral method [12]
can be used to show that in the limit $N \to \infty$ the free
energy, $f$, in $Z=\exp(-Nf)$, can be obtained as

$$ f=\min_x \left[ {kT x^2 \over 2 J} -Q(x) \right] \;\; ,
\eqno(2.4) $$

\NI where

$$ Q(x)=\ln \int  e^{x\sigma} d \mu (\sigma)\;\; . \eqno(2.5)$$

\NI{}If the minimum in (2.4) is at some $x=x_m$, then one can further
show that

$$ m=\left( {d Q \over d x} \right)_{x=x_m} =
{kT x_m \over J} \;\; , \eqno(2.6)$$

\NI where the last equality follows from the fact that the
global minimum is obtained at one (or more) roots of

$$ {d Q \over d x} = {kT x \over J} \;\; . \eqno(2.7)$$

\NI Thus, we note that $m=kTx_m/J$, i.e., $m\propto T$ for low
temperatures. This is one of those artificial infinite-range model
{}features. It turns out convenient to work with $x_m$ directly
rather than with $m$, as the order-parameter-like quantity for
series analysis. Of course, the actual
critical-tricritical-first-order behavior is at $T>0$ so the
difference only affects the form of analytic corrections to scaling.

In order to have a solvable model with tricritical behavior, we
take $Q(x)$ as an even, six-degree polynomial in $x$. There is still
{}freedom in selecting the coefficients, etc. We choose to
work with dimensionless parameters which, disregarding various
dimensional factors, amounts to effectively putting

$$ J=k/2 \;\; , \eqno(2.8)$$

$$ Q(x)=x^2+(U-1)x^4-x^6 \;\;  , \eqno(2.9)$$

\NI so that our choice corresponds to

$$ f=\min_x \left[ (T-1)x^2-(U-1)x^4+x^6 \right] \;\; . \eqno(2.10)$$

\NI This is the simplest Landau-expanded  form to yield the
tricritical point, at $(T,U)=(1,1)$, in the two-parameter space of
the (dimensionless) temperature, $T$, and another
(dimensionless) ``coupling constant,'' $U$.

On the low-$T$ side there are two symmetric roots of (2.7), and there
is always one root at $x=0$. We consider the root $x_m \geq 0$; the
\UN{\it actual series}\ is conveniently generated for

$$ x_m \sqrt{3} = \sqrt{ \sqrt{ (U-1)^2-3(T-1) } +U -1 }
\;\; . \eqno(2.11)$$

\NI By utilizing Mathematica, we derived the order 50 double series
in $T$ and $U$ for this order-parameter quantity. This series, i.e.,
the first 2601 coefficients $c_{ij}$, for $i,j=0, \ldots , 50 $, in

$$ \sqrt{3}\, x_m = \sum_{i=0} \sum_{j=0} c_{ij} T^i U^j \;\; ,
\eqno(2.12)$$

\NI{}can be obtained via electronic mail, from the authors, on request.

We also derived the functions $a_i (U)$ in

$$ \sqrt{3}\, x_m = \sum_{i=0} a_i (U) T^i \;\; ,
\eqno(2.13)$$

\NI{}for $i=0,1,\ldots,50$. These functions are available in
the FORTRAN form, via electronic mail.

\NP

\NI{\bf 3.\ \ THE TRICRITICAL PHASE DIAGRAM}

\

{}For $U<1$, there is a second-order transition line at $T=1$, at
which the order parameter approaches zero according to the
mean-field law $x_m \propto \sqrt{1-T}$. The proportionality
constant diverges as $1/\sqrt{1-U}$ for $U \to 1^-$. At the
tricritical point, the order-parameter vanishes according to
$\sim (1-T)^{1/4}$.

{}For $U>1$ there is a first-order transition. The line $T=1$ still
have special significance as the mean-field spinodal (see further
below) above which the high-temperature, zero-order-parameter phase
exists. However, it is not seen in the low-$T$ expansion. The actual
{}first-order transition line is determined by the condition that the
minima at $x_m > 0$ and at $x=0$ are equal; cf.~(2.10). A somewhat
lengthy calculation yields

$$ T=1+{1 \over 4}(U-1)^2 \eqno(3.1)$$

\NI for the first-order transition line at $U>1$.

Along this line
the low-$T$-side order parameter vanishes according to $\propto
\sqrt{U-1}$ as $U \to 1^+$, i.e., on approach to the tricritical
point. However, for fixed $U>1$, the order parameter is finite at the
{}first-order transition,

$$ x_m \sqrt{3} = \sqrt{3 \over 2} \sqrt{U-1} \;\; ,  \eqno(3.2)$$

\NI from the low-$T$ side, and it vanishes from the high-$T$ side.

In short-range Ising-type lattice models, if one attempts to
analytically continue the thermodynamic functions ``through'' the
{}first-order line, one encounters an essential singularity \UN{\it
at}\  the first-order transition. This singularity is due to droplet
excitations; it is weak and its detection in series analysis
has rarely been accomplished unambiguously [9]. Specifically, its
manifestation within the traditional Pad\'e method aimed at detecting
power-law divergences, is at best indirect via sequences of weak,
alternating poles and zeroes of the approximants [9]; see
the following sections for further discussion. While the incorporation
of this essential singularity must be ultimately a goal for a fully
systematic series-analysis method of tricriticality, at the present
state of the art and available series lengths, its presence will have
little effect in any series study.

There is no essential singularity for infinite-range models as there
are no droplet excitations, only uniform ones. When the thermodynamic
quantities are continued past the first-order transition, one
encounters a spinodal line at which the low-$T$ order parameter $x_m
>0$ ceases to be a local minimum of the constrained free energy.
(While at the first-order line it ceases to be the global minimum.)
This mean-field spinodal line is at

$$ T=1+{1 \over 3}(U-1)^2 \;\; . \eqno(3.3)$$

\NI The existence of a sharp spinodal-type singularity is an
artifact of the infinite-range model. However, for short-range models
traces of spinodal-type behavior have been noted in available-length
series analyses [9]. These are artifacts of employing
approximants (within Pad\'e or other analysis methods) which fit the
data to a form suggestive of a sharp singularity; see [9] for
{}further discussion.

Near the tricritical point, one can write the low-$T$-side scaling
{}form in terms of the scaling variables

$$ t=T-1 < 0 \qquad {\rm and} \qquad u=U-1 \;\; , \eqno(3.4)$$

$$ \sqrt{3} x_m \simeq (-t)^{1/4} F_- \left( u (-t)^{-1/2} \right)
\;\; , \eqno(3.5)$$

\NI where the scaling form (3.5) applies for $t, u \to 0$ and
the scaling function is

$$ F_-(\zeta) = \sqrt{\zeta+\sqrt{\zeta^2+3 }} \;\; . \eqno(3.6)$$

\NP

\NI{\bf 4.\ \ SLICEWISE PAD\'E ANALYSIS}

\

The slicewise Pad\'e analysis is perhaps the simplest, single-variable
approach to analyzing double-series expansions [7]. Thus, we
calculate, for fixed $U$, approximants to the series coefficients $a_i$
in

$$ \sqrt{3}\, x_m(T,U{\rm -fixed})=\sum_{i=0} a_i (U) T^i \;\; .
\eqno(4.1)$$

\NI The coefficients $a_i$ are approximated by

$$ a_i \simeq \sum_{j=0}^{j_{max}} c_{ij} U^j \;\; , \eqno(4.2)$$

\NI where in our case $j_{max}=50$.

The order $[M/L]$ dlog-Pad\'e approximant to the derivative $x_m^\prime =
\partial x_m / \partial T$ is defined as the rational approximant of
the form

$$ {x_m^{\prime\prime} \over x_m^\prime} \simeq
{ p_0 + p_1 T + p_2 T^2 + \ldots + p_M T^M \over
1 + q_1 T + q_2 T^2 + \ldots + q_L T^L } \;\; . \eqno(4.3)$$

\NI where the derivative $x_m^\prime$ rather than $x_m$ was used in
order to have both the numerator and \UN{\it denominator} of the
left-hand side diverge at their first singularity as $T$ is
increased from zero. Specifically, for fixed $U$, we have

$$ x_m^\prime \propto [T(U) - T]^{-B} \;\; , \eqno(4.4)$$

\NI with $T(U)=T_c=1, B=1/2$ at the critical line for $U<1$,
and $T(1)=1, B=3/4$ at the tricritical value $U=1$. For the
infinite-range model $T(U)$ equals the spinodal value (3.3), while
$B=1/2$, for fixed $U>1$. Note that for $U \leq 1$ the exponent $B$ is
related to the order-parameter exponent usually denoted by $\beta$, via
$B=1-\beta$.

The coefficients $p_{0, \ldots, M}$ and $q_{1, \ldots, M}$ are
calculated in a standard fashion [13] to have the power series of
the right-hand side of (4.3) reproduce the first $M+L+1$ power-series
coefficients of the left-hand side. Of special interest are the poles
of the approximant which here depend parametrically on $U$ and will
be loosely denoted simply by $T(U)$. It is anticipated that for
power-law singularities (4.4) a ``stable'' pole location will be found
in the highest-order, near diagonal approximants (i.e., $M \simeq L$
and $M+L+1$ close or equal to the order of the available series for
$x_m^{\prime\prime} / x_m^\prime$) such that for $T$ near $T(U)$ the
right-hand side of (4.3) approximates the behavior suggested by (4.4),

$$ {x_m^{\prime\prime} \over x_m^\prime} \simeq {B \over T(U)-T} \;\; .
\eqno(4.5)$$

\NI Thus, the residue at the stable pole approximates the exponent
in (4.4), \ $-B$.

{}For essential singularities and other singularities
associated with branch cuts, it has been noted [9] that Pad\'e
approximants sometimes yield a sequence of alternating weak poles and
zeros (zeros of the denominator and numerator, respectively) which
mimic the branch cut. However, the Pad\'e method is well suited only
{}for single-variable expansions of functions with power-law
singularities. It is worth pointing out that recently exact results
were derived for certain models of partially convex lattice vesicles
[14] which show tricritical behavior with essential singularities at
the ``first-order line'' of their phase diagram. Double-variable
expansions can be derived for these models [15] and possibly
used as test series for methods to detect essential singularities, etc.
However, we note that the singularities of the models of [14,15] seem to
be natural-boundary-type and differ from droplet-type essential
singularities anticipated at Ising first-order transitions.

It is also important to point out that in the slicewise
Pad\'e method used here the approximation if two-fold: the
coefficients $a_i$ are calculated \UN{\it approximately}\ via the
truncated series (4.2); the Pad\'e method is applied to the truncated
series (4.1). We kept the order of the Pad\'e approximation,  $M+L+1$,
at about half the order of the truncation in (4.2), which is
$j_{max}=50$. Still, as examples below illustrate, the method fails
{}for $U^>_\sim 1$ which is presumably due to the truncation (4.2).
While the truncation (4.2) is the approach used in the early
literature, it is natural to consider improvement of this
approximation: we address this issue later in this section.

As our first example, it is useful to consider approximants with
only a single pole. For instance, Figure~1 shows the pole location for
the approximant
$[19/1]$. We note that the approximant becomes ``defective'' for
$U^>_\sim 1$ in that the location of the pole has nothing to do with
the actual spinodal line. Figure~2 also illustrates that the residue
provides a poor approximation to the exponent $B$. The latter,
however, can be blamed on our use of the extremely off-diagonal
approximant. This ``hooking'' of the approximant away from the actual
phase-transition line has been noted in earlier studies [7]. The
hooking can also be in the direction opposite to that of Figure~1, as
occurs, for instance, in the $[30/1]$ approximant not shown here.

Study of diagonal and also numerous near-diagonal approximants (only
two are actually illustrated here; see Figures~3, 4, 5) reveals that
they indeed significantly improve the exponent $B$ estimate for small
$U<1$. Most approximants also yield $B$ values quite close to $3/4$ at
$U=1$. However, the phase diagram is not well-represented near and
above the tricritical point. In the wide crossover regime near $U=1$,
the approximants become defective: the ``physical'' pole alternates
among several branches of the roots of the denominator; see Figure~3
{}for the $[10/10]$ approximant. While these branches first roughly follow
the exact spinodal location above $U=1$, they soon ``hook away.'' The
exponent estimates also show a rather irregular crossover pattern near
$U=1$; see for instance Figure~4 for $[10/10]$. The approximation fully
deteriorates soon above $U=1$. Furthermore, it seems that the quality
of the approximation is affected little by increasing the approximant
order as illustrated by the case $[17/17]$ in Figure~5.

Our conclusion is that the slicewise Pad\'e method as implemented [7],
can at best provide a qualitative indication of the
presence of a tricritical point. A wide region of irregular,
``defective'' approximant behavior develops on approach to
tricriticality. The hooking noted in earlier studies [7] can be
attributed to changeovers among the various pole branches. Thus
plotting all the poles and all the residues allows a rough location
of the tricritical point and estimation of the exponent. But
otherwise the slicewise Pad\'e method should not be regarded as a
systematic technique for analyzing tricritical behavior.

We now turn to the approximation involved in the truncation (4.2). Some
series actually have the functions $a_i(U)$ as polynomials in $U$ (or
other appropriate expansion parameter). However, generally the
double-series are available in the form truncated in both variables.
It is natural to assume that the quality of the overall approximation
can be improved by using resummation methods for the single-variable
series (4.2). To our knowledge, no systematic procedures were developed
in the literature; one could contemplate Pad\'e or other resummation
methods. However, in this work we limited ourselves to the following: we
repeated the preceding analysis with the \UN{\it exact}\  functions
$a_i(U)$, derivation of which was described in Section~2.

{}Figure~6 illustrates the behavior of the poles for the case of the
[10/10] Pad\'e approximant. It should be compared with Figure~3. We
note that without the truncation (4.2), the approximation for $U^>_\sim
1$ improves. Specifically, the leading pole now clearly follows the
spinodal line for $U>1$. However, the main problems remain: the
crossover region near $U=1$ is marked by the ``defective'' behavior
where poles approach each other and switch role. Furthermore, the
leading pole for $U>1$ is accompanied by a sequence of weak poles which
mimic branch-cut effect. Figure~7 shows the residues at the poles of
the [10/10] Pad\'e approximant. It should be compared with Figure~4.
The exponent estimation is only
accurate for $U<1$. In the wide crossover region near $U=1$ the
defective behavior spoils the accuracy of the approximation, while for
$U>1$, the quality of the approximation is low presumably due to the
accompanying weak poles. A similar behavior was observed for other
near-diagonal Pad\'e approximants calculated with the exact $a_i(U)$
values, and as before there was no visible improvement when the order,
$[M/L]$, of the approximant was increased.

\NP

\NI{\bf 5.\ \ CONCLUDING REMARKS}

\

As discussed in the previous section, the simplest Pad\'e approach
{}fails in several aspects near tricritical points. Let us now consider
a ``wish list'' for a more systematic series analysis method. Firstly,
we would expect to produce a smoothed out but regular approximation to
the exponent, i.e., the spiked line in Figure~2, and to the phase
diagram, i.e., to the phase-transition lines shown in Figure~1. The
approximants should be sharpening up with the increased order of
approximation. Secondly, we would also like to estimate the scaling
{}form at the tricritical point, cf.~(3.5), as in the
differential-approximant analyses of bicritical points [11].

An added complication at tricritical points is the presence of
singularities at the first-order transition line, as well as
possible pseudo-singular spinodal behavior. Specifically, for
Ising-type models there are weak essential singularities.
However, for other models with soft-mode excitations, power-law
spin-wave-type singularities are present at the first-order transition.
{}For yet another class of models, the droplet picture may
not be fully understood, such as for certain systems with randomness,
or the first-order-regime singularities may not have been carefully
discussed in the literature, such as for the polymer collapse.

We note that the deterioration of the approximant quality in the
slicewise Pad\'e method has always occurred at larger $U$ values.
Improvement accomplished by avoiding the approximate truncation
(4.2) was not sufficient for really accurate results. Other
approaches involve, for instance, ``slicing'' along
curvilinear paths that originate at the origin of the $(T,U)$
plane, etc. These possibilities will be explored in future publications.

All the above remarks indicate that series analysis of tricritical
behavior promises to become an interesting and active field
with the availability of new, long series and modern computational
{}facilities. It is hoped that the approach presented here will yield
useful test series for these studies.

JA acknowledges support from the US--Israel Binational Science Foundation.
VP wishes to acknowledge the
hospitality and financial assistance of the Institute for Theoretical
Physics at the Technion --- Israel Institute of Technology, where this
work was initiated.

\NP

\centerline{\bf REFERENCES}{\frenchspacing

\

\item{[1]} A. Aharony, Phys. Rev. B{\bf 18}, 3318 (1978); for
recent series analysis studies of random-field models, see M. Gofman,
J. Adler, A. Aharony, A.B. Harris and M. Schwartz, Phys. Rev. Lett.
{\bf 71}, 1569 (1993).

\item{[2]} T. Ishinabe, J. Phys. A{\bf 18}, 3181 (1985);
V. Privman, J. Phys. A{\bf 19}, 3287 (1986).

\item{[3]} K. Binder and D.P. Landau, Phys. Rev. B{\bf 21}, 1941 (1980).

\item{[4]} I. Chang and H. Meirovitch,
Phys. Rev. E{\bf 48}, 3656 (1993), and references therein;
H. Meirovitch and H.A. Lim,
J. Chem. Phys. {\bf 92}, 5155 (1990) and
{\bf 91}, 2544 (1989).

\item{[5]} J.D. Kimel, P.A. Rikvold and Y.-L. Wang,
Phys. Rev. B{\bf 45}, 7237 (1992);
B. D\"unweg, A. Milchev and P.A. Rikvold,
J. Chem. Phys. {\bf 94}, 3958 (1991).

\item{[6]} C.C.A. G\"unther, P.A. Rikvold and M.A. Novotny,
Phys. Rev. B{\bf 42}, 10738 (1990), and references therein;
J.B. Collins, P.A. Rikvold and E.T. Gawlinski,
Phys. Rev. B{\bf 38}, 6741 (1988).

\item{[7]} J. Oitmaa,
J. Phys. C{\bf 5}, 435 (1972) and
C{\bf 4}, 2466 (1971);
{}F. Harbus and H.E. Stanley,
Phys. Rev. B{\bf 8}, 1141 and 1156 (1973).

\item{[8]} W. Janke, Phys. Rev. B{\bf 47}, 14757 (1992).

\item{[9]} V. Privman and L.S. Schulman, J. Stat. Phys. {\bf 29}, 205
(1982);
J. Adler and D. Stauffer, Physica A{\bf 175}, 222 (1991);
see also
R.V. Ditzian and L.P. Kadanoff, J. Phys. A{\bf 12}, L229 (1979).

\item{[10]} D.M. Saul, M. Wortis and D. Stauffer, Phys. Rev. B{\bf 9},
4964 (1974). Recently, a similar approach was used to locate first-order
transitions in certain 2D and 3D Potts models: K.M. Briggs, I.G. Enting
and A.J. Guttmann, J. Phys. A (1994), in print; A.J. Guttmann and I.G.
Enting, J. Phys. A (1994), in print. Earlier attempts along these lines
were reported in [7,9].

\item{[11]} M.E. Fisher, J.-H. Chen and H. Au-Yang,
J. Phys. C{\bf 13}, L459 (1980).

\item{[12]} M.L. Glasser, V. Privman and L.S. Schulman,
Phys. Rev. B{\bf 35}, 1841 (1987), see Section~IV there.

\item{[13]} G.A. Baker, Jr., {\it Qualitative Theory of Critical
Phenomena}, Academic Press, San Diego (1990), see Part~III there.

\item{[14]} Review: T. Prellberg and A.L. Owczarek, Mod. Phys. Lett. A
(1994), in print.

\item{[15]} A.J. Guttmann, private communication (1994). Details of the
singularity structure have been elucidated by T. Prellberg and R. Brak,
J. Stat. Phys. (1994), in print.

}\NP

\centerline{\bf FIGURE CAPTIONS}

\

\NI\hang{}{\bf Figure~1:}\ \ The symbols $\diamond$, merging for most
part as heavy solid lines, show the pole $T(U)$ of
the $[19/1]$ dlog-Pad\'e approximant to the $T$-derivative of the
magnetization series. The thin solid lines correspond to the
{}first-order transition (lower curve, equation (3.1)), and to the
mean-field spinodal line (upper curve, equation (3.3)), for $U >
1$. The tricritical point is at $T=1, U=1$, while for $U<1$ there is
the second-order transition at $T \equiv 1$.

\NI\hang{}{\bf Figure~2:}\ \ The symbols $\diamond$ show the exponent
$B(U)$ as calculated from the residue at the pole of
the $[19/1]$ Pad\'e approximant. The horizontal thin solid line
corresponds to the exact value $B=1/2$ for $U \neq 1$. The
vertical spike at $U=1$ goes up to the exact value $B(1)=3/4$.

\NI\hang{}{\bf Figure~3:}\ \ Shown are all the poles $T(U)$ of the
$[10/10]$ Pad\'e approximant which fit within the figure range.

\NI\hang{}{\bf Figure~4:}\ \ Exponent estimates $B(U)$ from the
residues of all the poles of the $[10/10]$ Pad\'e approximant.

\NI\hang{}{\bf Figure~5:}\ \ Shown are all the poles $T(U)$ of the
$[17/17]$ Pad\'e approximant.

\NI\hang{}{\bf Figure~6:}\ \ Shown are the poles $T(U)$ of
the $[10/10]$ dlog-Pad\'e approximant calculated with the exact
coefficient values $a_i(U)$; cf.~Figure~3.
{}For $U>1$, the thin solid line corresponds to the
{}first-order transition (equation (3.1)), while the
leading approximant (the smallest $T(U)$ values) follows the
mean-field spinodal (equation (3.3)), with the difference smaller
than the size of the symbols.

\NI\hang{}{\bf Figure~7:}\ \ The exponent
$B(U)$ as calculated from the residue at the pole of
the $[10/10]$ Pad\'e approximant with the exact $a_i(U)$ values;
cf.~Figure~4. The horizontal thin solid line
corresponds to the exact value $B=1/2$ for $U \neq 1$. The
vertical spike at $U=1$ goes up to the exact value $B(1)=3/4$.

\bye